\documentclass[a4paper]{jpconf}
\usepackage{graphicx}

\usepackage{amsmath}
\usepackage{amssymb}
\usepackage{graphicx}

\def\Journal#1#2#3#4{#4 \textit{#1} {\bf #2} #3 }

\def\PLB{Phys. Lett. B}

\def\PRL{Phys. Rev. Lett.}

\def\PRC{Phys. Rev. C}

\def\RMP{Rev. Mod. Phys.}

\DeclareMathAlphabet{\mathcmsy}{OMS}{cmsy}{m}{n}
\SetMathAlphabet{\mathcal}{normal}{OMS}{mdbch}{m}{n}

\newcommand{\braket}[2]{\ensuremath{\langle{#1}|{#2}\rangle}}
\newcommand{\onebody}[3]{\ensuremath{\langle{#1}|{#2}|{#3}\rangle}}

\newcommand{\bvec}[1]{\ensuremath{\boldsymbol{#1}}}

\newcommand{\tgenm}[1]{\ensuremath{\boldsymbol{\mathcal{#1}}}}

\newcommand{\matnot}[1]{\ensuremath{\mathbf{#1}}}

\newcommand{\op}[1]{\ensuremath{\Hat{\mathcmsy{#1}}}}

\newcommand{\lat}[1]{\ensuremath{\mathfrak{#1}}}

\newcommand{\re}{\operatorname{\Re e}}

\begin{document}
\title{Fully antisymmetrised dynamics for bulk fermion systems}

\author{Klaas Vantournhout and Hans Feldmeier}
\address{GSI Helmholtzzentrum f\"{u}r Schwerionenforschung GmbH\\
Planckstra\ss e 1\\
64291 Darmstadt, Germany}
\ead{k.vantournhout@gsi.de}

\begin{abstract}
The neutron star's crust and mantel are typical examples of non-uniform bulk systems with spacial localisations. When modelling such systems at low temperatures, as is the case in the crust, one has to work with antisymmetrised many-body states to get the correct fermion behaviour. Fermionic molecular dynamics, which works with an antisymmetrised product of localised wave packets, should be an appropriate choice. Implementing periodic boundary conditions into the fermionic molecular dynamics formalism would allow the study of the neutron star's crust as a bulk quantum system. Unfortunately, the antisymmetrisation is a non-local entanglement which reaches far out of the periodically repeated unit cell. In this proceeding, we give a brief overview how periodic boundary conditions and fermionic molecular dynamics can be combined without truncating the long-range many-body correlation induced by the antisymmetry of the many-body state.
\end{abstract}

\section{Introduction}
Neutron stars represent one of the densest forms of matter in the observable universe. As an ``outgrown nucleus'', they confine approximately $10^{57}$ baryons inside a radius of roughly 10 km. This leads, in the centre, to densities expected to be as large as five to ten times the normal nuclear saturation density $n_0$. In the outer layers of a neutron star, the outer and inner crust which represents only a tiny fraction of the total mass, the densities are on the average below $n_0$. Nonetheless, it is associated with various astrophysical phenomena such as the cooling of neutron stars or neutron star glitches \cite{chamel2008}. A detailed knowledge of crustal matter will help to improve our understanding of these phenomena.\\

The neutron star's crust consists of protons, neutrons and electrons that organise themselves as a result of the combined effect of the short-range nuclear attraction and the long-range Coulomb repulsion. In the outer crust, i.e.\ the lower-density regions, nucleons bind to form nuclei that are placed in a Coulomb lattice embedded in an electron gas. In the inner crust, the neutron-drip density ($\approx4 \times 10^{11} \mathrm{g}/\mathrm{cm}^3$ \cite{ruster2006}) is surpassed giving rise to neutron rich nuclei placed on a similar Coulomb lattice, immersed in a gas of neutrons and electrons. In the more dense layers of the crust ($n_0/3$ to $n_0/2$ \cite{haensel2007}), the spherical nuclei outgrow the lattice spacing. In this crust-core interface, i.e.\ the mantle of the neutron star, it is believed that the system balances on a subtle interplay between the nuclear surface energy and the Coulomb energy of the neutron-proton-electron system. This results in a multitude of competing quasi-ground states with similar energies, often having quite different matter distributions where the system can cool down to. These complex-shaped foam like structures are referred to as ``nuclear pastas'' \cite{chamel2008, ravenhall1983, sonoda2008}.\\

\section{Fermionic molecular dynamics}
The crusts and especially the mantel of a neutron star are excellent examples of bulk fermion matter where spatial localisations play an important role. Many computational techniques have been used to investigate these. These range from the liquid drop model over Hartree Fock to Monte Carlo and molecular dynamics \cite{ravenhall1983, oyamatsu1984, horowitz2004, watanabe2009, nakazato2009, newton2009, avancini2010}. With a few exceptions, all of these methods study a ``single'' geometry of the nuclear cluster. This geometry is either explicitly implemented or strongly correlated to the shape of the unit cell and the imposed boundary conditions. However, the implementational or computational constraints should not distort the properties of the macroscopic system. The Monte Carlo and molecular dynamics methods are unbiased with regards to the geometry of the nuclear clusters and allow the study of more realistic systems. As there are many configurations with almost the same energy, the nuclear pasta phases are susceptible to small perturbations stemming from external probes or slight temperature changes. The molecular dynamics picture is appropriate to study such systems and is already successfully used for classical descriptions. From the many existing approaches  \cite{allen1991, aichelin1986, maruyama1996, feldmeier2000, ono2004}, classical molecular dynamics (CMD) provided a way to study the effect of large density fluctuations on neutrino opacities and evaluated the breaking strain of the neutron star's crust \cite{horowitz2004, horowitz2009}. Quantum molecular dynamics (QMD), on the other hand, identified various pasta structures and evaluated one of the first phase diagrams of the different pasta phases \cite{watanabe2009}.\\

Disregarding the classical nature of CMD and QMD, their achieved successes in the study of nuclear matter at sub-nuclear densities have given us a first understanding of the crust's bulk properties. However, when studying phenomena at length scales smaller then the de Broglie wavelength of the system's constituents, quantum effects stemming from the wave character of the nucleons predominate. In addition, the statistical many-body correlations related to the system's fermionic content will play a fundamental role. While CMD lacks both the wave character as well as the statistical description, QMD works with density packets representing distinguishable particles. The indistinguishablity of the fermions is mimicked by means of a phenomenological Pauli potential which prevents identical fermions to occupy the same phase space \cite{dorso1987}. Another fundamental quantum feature missing in QMD is the dynamical variation of individual spin components and the spreading of the wave packets.\\

The shortcomings of QMD for fermion systems are remedied in fermionic molecular dynamics (FMD) \cite{feldmeier2000, ono2004}. While QMD uses a phenomenological Pauli potential to mimic fermion statistics, FMD introduces the fermion many-body states as antisymmetrised products, i.e.\ Slater determinants, of localised single particle states $|q_p(\bvec z_p)\rangle$\,:
%---------------------------------------------------------------------
\begin{equation}\label{eq:many-body}
|Q(\bvec z_1,\ldots,\bvec z_A)\rangle = \op{A}|q_1(\bvec z_1)\rangle\otimes\cdots\otimes|q_A(\bvec z_A)\rangle.
\end{equation}
%---------------------------------------------------------------------
This way, the Pauli exclusion principle and uncertainty principle are implemented from the beginning and cannot be violated. With a proper choice of the single-particle states $|q_p(\bvec z_p)\rangle$ and their variation parameters $\bvec z_p$, concepts as average phase-space position, spin, isospin and spreading can be introduced. While the equations of motion of the many-body fermion system can be obtained by means of the time-dependent variational principle, structure studies are obtained by means of Ritz' variational principle \cite{neff2008}.\\

When studying the observable properties of fermion systems represented by many-body states as Eq.~\eqref{eq:many-body}, one has to evaluate the expectation values of various operators. Although the single-particle states in the FMD wave function are not orthonormal, the expectation values can easily be evaluated by means of a matrix formalism as
%---------------------------------------------------------------------
\begin{equation}\label{eq:finite:operators}
\mathcal{B}_I =
\sum_{pq=1}^A\onebody{q_p}{\op{B}_{I}}{q_q}\matnot{o}_{qp},\qquad
\mathcal{B}_{II}=
\frac{1}{2}\sum_{pqrs=1}^A \onebody{q_pq_r}{\op{B}_{II}}{q_qq_s}
(\matnot{o}_{qp}\matnot{o}_{sr} -
\matnot{o}_{qr}\matnot{o}_{sp}).
\end{equation}
%---------------------------------------------------------------------
The wave function's determinant structure is embodied in the matrix $\matnot o$, representing the inverse of the overlap matrix $\matnot n$ with $\matnot n_{pq} = \braket{q_p}{q_q}$. These two matrices are fundamental for a correct fermion description. Therefore, and because their eigenvalues can cover many orders of magnitudes, it is paramount to calculate these matrices as accurately as possible (i.e.\ analytically).\\

FMD was originally devised to describe heave-ion reactions in a time-dependent framework \cite{feldmeier2000, ono2004}. Now, FMD is also successfully used to describe stationary situations related to nuclear structure and nuclear reactions by mixing many Slater determinants \cite{neff2008, neff2011}. However, due to the long-range character of the Pauli correlations, evaluating bulk fermion matter becomes a tedious task. The dimension of the matrices usually impedes simulations of a large number of particles, however desirable these may be for bulk fermion systems such as crustal matter. Applying FMD on bulk fermion systems has therefore been a long open question. This has recently been solved within the concept of ``periodic boundary conditions'' by making the spatial positioning of the single-particle states periodic \cite{vantournhout2011a, vantournhout2011b}. An infinite fermion system is created by replicating a unit cell---containing $A$ particles---on an infinite lattice \lat B.  Each cell can be identified by a lattice vector $\bvec R=n_1\bvec a_1 + n_2\bvec a_2 + n_3\bvec a_3$ with integer $n_j$. Using the translation operator $\op{T}(\bvec R)$, the FMD many-body state can be written as
%---------------------------------------------------------------------
\begin{equation}\label{eq:tapbc:W}
  |Q_\infty(\bvec z_1,\ldots,\bvec z_A)\rangle = \op{A}\bigotimes_{\bvec R\in\lat B} \op{T}(\bvec R)
  \left\{|q_1(\bvec z_1)\rangle\otimes\cdots\otimes|q_A(\bvec z_A)\rangle\right\}.
\end{equation}
%---------------------------------------------------------------------
As indicated earlier, the overlap matrix $\matnot n$ and its inverse $\matnot o$ are fundamental for the correct fermion description of the system. Although both matrices have infinite dimensions, they exhibit a peculiar nested block-Toeplitz structure which can be exploited \cite{vantournhout2011a}. As a result of the translation invariance of the system, each $A\times A$ block of the overlap matrix can be identified unambiguously with the lattice vector $\bvec R$ that connects the two cells of the bra and ket states. The blocks can be evaluated as $\matnot n_{pq,\bvec R} = \langle q_p|\op{T}(-\bvec R)|q_q\rangle$. Although $\matnot n$ and $\matnot o$ are of infinite dimension, it is possible to obtain the inverse overlap matrix through the following scheme
%---------------------------------------------------------------------
\begin{equation}
\tgenm N(\bvec k) = \sum_{\bvec R\in\lat B} \matnot n_{\bvec R}\,
  e^{-i\bvec k\cdot\bvec R},\qquad \tgenm O(\bvec k) = \tgenm
  N(\bvec k)^{-1}, \qquad\matnot o_{\bvec R} = \frac{1}{V_{BZ}}\int_{BZ}\tgenm O(\bvec
  k)\,e^{i\bvec k\cdot\bvec R}\,\textrm{d}^3\bvec k,
\end{equation}
%---------------------------------------------------------------------
where $BZ$ represents the first Brillouin zone of the lattice $\lat B$ and $\bvec k$ is a vector in this volume \cite{vantournhout2011a}.\\

The strength of the proposed formalism reveals itself upon evaluating the operators in reciprocal space.  Normally, the expectation values of the infinite fermion system would be calculated by means of Eqs.~\eqref{eq:finite:operators} resulting in infinite sums over the block structure. These sums, however, translate into integrals over the first Brillouin zone which are computationally straightforward to evaluate. The expectation value of a one-body operator per unit-cell volume can then be computed as
%---------------------------------------------------------------------
\begin{equation}\label{eq:pbc:operators:bloch}
    \mathcal{B}_{\rho,I} = \frac{1}{V_{BZ}}\int_{BZ} \sum_{pq=1}^A\tgenm
    B_{I,pq}(\bvec k)\tgenm O_{qp}(\bvec k)\,\textrm{d}^3\bvec k ,\qquad
    \tgenm B_{I,pq}(\bvec k) = \sum_{\bvec R\in\lat B}
    \onebody{q_p}{\op{B}_{I}\op{T}(\bvec R)}{q_q}\,
    e^{i\bvec k\cdot\bvec R}.
\end{equation}
%---------------------------------------------------------------------
Here we assume that the operator $\op{B}_I$ commutes with the translation operator $\op{T}(\bvec R)$ over any lattice vector $\bvec R$. The two-body operator has a similar structure.

\section{Results}
When studying properties of bulk matter, it is important that the constraints of the simulation do not interfere with the result and that the properties of the simulation should be equivalent with those of the macroscopic system. It stands to reason that for non-periodic systems, the periodicity should not influence the simulation's results. As long as the correlation volume of the interaction does not exceed the simulation volume, the imposed periodicity works fine. However, serious problems arise in the presence of long-range correlations such as induced by the Pauli exclusion principle. We demonstrate that the proposed technique reproduces two essential features intrinsic to free fermions. The single-particle states are Gaussian wave packets of the form $\langle \bvec x|a\bvec b\rangle = \exp\{-(\bvec x-\bvec b)^2/(2a)\}$ where the complex vector $\bvec b$ represents the mean phase-space position and $a$ is a complex parameter connected with the spreading of the wave packet in phase space.\\

\begin{figure}[b]
\begin{center}
\includegraphics[width=\textwidth]{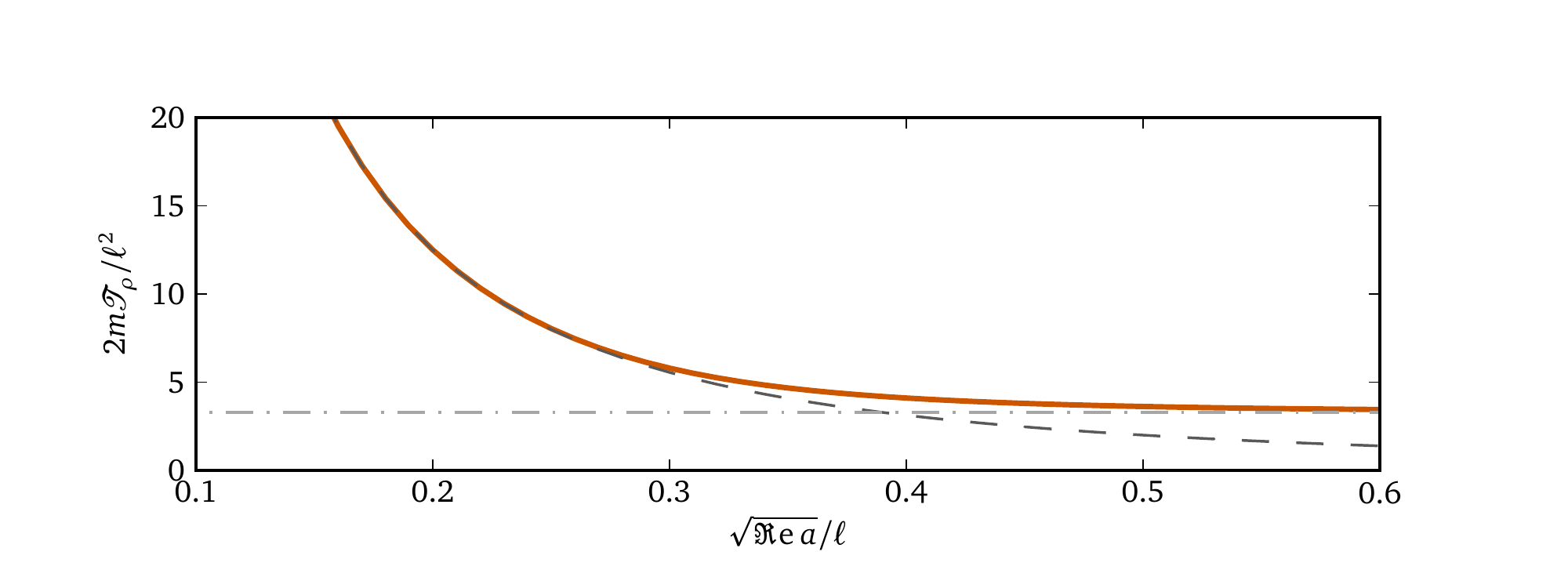}
\caption{The kinetic energy of periodic Gaussian wave packets (solid line) is compared with that of a single Gaussian (dashed line) and the free Fermi gas (dash-doted line). The kinetic energy per particle $\mathcal T_\rho$ is scaled with the mass $m$ of the fermion and the periodic spacing $\ell$ (box-size) of the Gaussians. The kinetic energy is presented as a function of the scaled variance $\sqrt{\re a}/\ell$ of the Gaussian wave packets.}\label{fig:ekin}
\end{center}
\end{figure}

In Ref.~\cite{feldmeier2000} it was shown that, for a hundred periodically positioned wave packets, Eqs.~\eqref{eq:finite:operators} reproduce the momentum and spatial densities intrinsic to a one-dimensional Fermi system. An identical result was achieved in Ref.~\cite{vantournhout2011a} by using periodic boundary conditions with a unit cell containing a single particle. The densities were evaluated using Eqs.~\eqref{eq:pbc:operators:bloch}. In this reference, it was shown that with increasing overlap of neighbouring wave packets, the effect of antisymmetrisation makes the fermionic behaviour apparent. This effect is also seen in Fig.\ \ref{fig:ekin}
%---------------------------------------------------------------------
%---------------------------------------------------------------------
which compares the kinetic energy of the same periodic system with that of an individual Gaussian and that of a free one-dimensional Fermi gas as function of the wave packet spreading. Although this one-dimensional case hints towards free fermion behaviour, it must be mentioned that, in one dimension, the free Fermi gas results and that of a weak periodic system coincide when having identical particle densities. In more dimensions, it can be shown that the one-particle-in-a-box results evolve towards those of a weak periodic system \cite{vantournhout2011a}. This effect is expected because with increasing overlap, ergo increasing wave packet spreading, the wave packet seems uniform to the unit cell and the system behaves as if it consisted of plane waves in a weak periodic system. A successful investigation of bulk matter obviously requires that influences of the geometry of the chosen boundary conditions are negligible, and thus the wave packets should be smaller then the unit cell.\\

\begin{figure}[t]
\begin{center}
\includegraphics[width=0.45\textwidth]{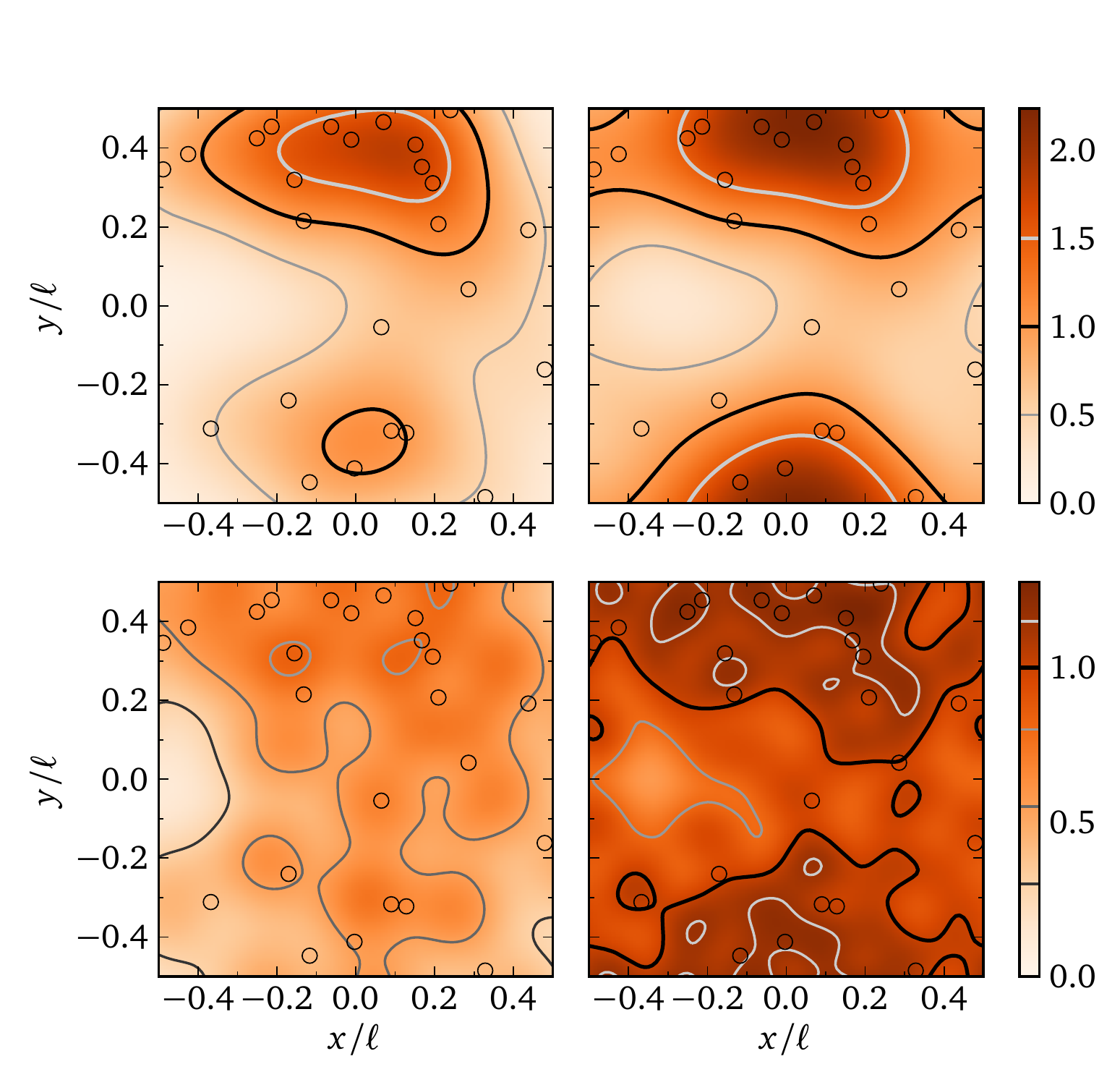}
\includegraphics[width=0.45\textwidth]{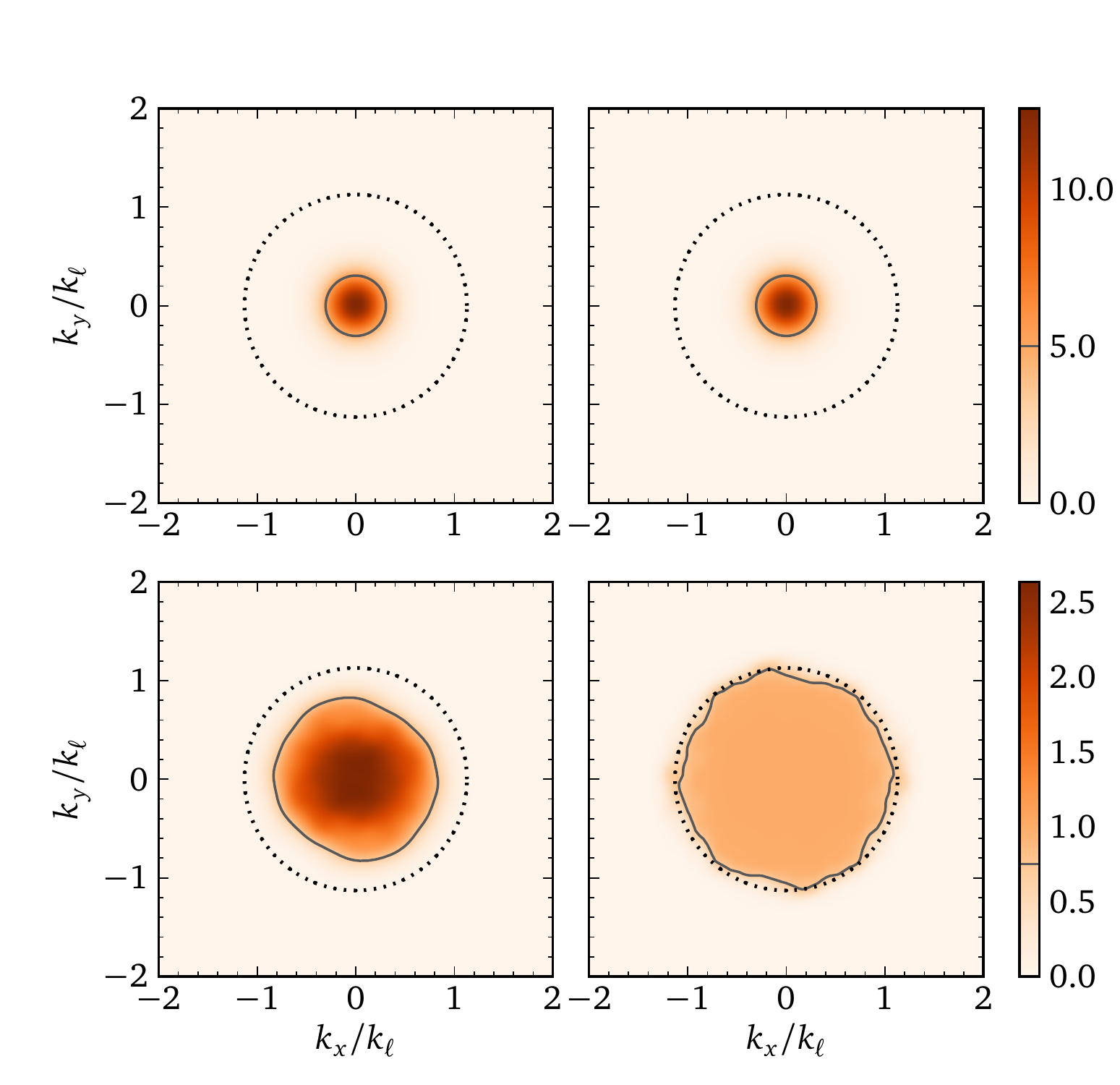}
\caption{Comparison of the scaled density distributions (left four panels\,: spatial $\rho_x \ell^2/N$; right four panels\,: momentum $\rho_k k_\ell^2/N$) of $N=25$ Gaussian wave packets under different conditions. The wave packets are randomly placed in the unit cell of a square lattice with unit length $\ell$ and have a width parameter given by $a=(0.2\ell)^2$. The mean momentum of the individual wave packets is set to zero. The spatial densities are normalised and coordinates are expressed in units of $\ell$. Analogously for the momentum density with coordinates expressed in units of $k_\ell = \sqrt{N}\pi/\ell$. For each type of density distribution\,: the upper-two panels depict distinguishable particles, the lower-two panels represent the antisymmetrised case, the left and right panels show the effect of a calculation without and with periodicity, respectively, as described in the text. For the spatial density, the black circles show the positions of the centroids of the wave packets. For the momentum density, the dotted circle represents the Fermi ``sphere'' of a system of free particles with the same mean density, $k_F = \sqrt{4\pi N}/\ell$. Note the difference in colour scaling.}\label{fig:densdistr}
\end{center}
\end{figure}

Fig.\ \ref{fig:densdistr}
%---------------------------------------------------------------------
%---------------------------------------------------------------------
depicts the effect of periodicity and antisymmetry on the spatial (left hand side) and momentum distributions (right hand side) for 25 single-particle states with zero mean momentum randomly distributed in a square unit cell. For distinguishable particles (upper four panels) the momentum distribution of the 25 particles is the same as the one given by the individual wave packet. When enabling the periodic boundary conditions, only the spatial density grows at the boarders due to the tails leaking in from the neighbouring cells. For indistinguishable fermions (lower four panels), where the many-body wave function is antisymmetrised, one sees that the spatial density (without periodic boundary conditions) in the lower left most panel is much smoother. The fermions tend to avoid closeness in coordinate space and at the same time enlarge the occupied momentum space as indicated in the corresponding panel for the momentum density, thus, inducing Fermi motion. When enabling the periodic boundary conditions, the spatial density increases again due to the tails of the neighbouring cells. In momentum space, the distribution is even more spread due to the antisymmetrisation which is now also with respect to the particles in all surrounding cells, neighbouring and beyond. One obtains an almost uniform Fermi distribution where deformations in the surface reflect the clustering of the spatial density. Hence, a simulation of clustered densities with the proposed periodic structure reproduces intrinsic bulk fermion behaviour, unaffected by the imposed periodic structure which leaves no trace in the momentum distribution. This clearly shows that antisymmetry, a long-range many-body correlation, should not be ignored or truncated for the bulk description.

\section{Conclusion}
This proceedings gave a short overview how periodic boundary conditions can be implemented to study fully antisymmetrised, infinitely extended, inhomogeneous fermion matter by means of localised single-particle states. The bulk matter is created by spatially distributing a finite number of wave packets in a unit cell which is periodically repeated over space. By treating each single-particle state as an individual state, the overlap matrix and its inverse are of infinite size but possess a nested Toeplitz structure. This is a feature that can be exploited to reduce the computational cost and eliminate the infinite dimensionality of the equations. This allows us to study bulk fermion matter by means of a finite number of fermions. The structure of the resulting equations to evaluate expectation values, resembles those of a finite fermion system, but requires an extra integration over the first Brillouin zone. The later is a translation of the periodic boundary conditions. Although the equations only address a finite number of particles in a unit cell, they keep track of the Fermi statistics of the infinite system. It is demonstrated that this formalism incorporates intrinsic bulk fermion behaviour and reproduces density distributions and energies of free fermion systems. Furthermore, the use of localised states makes the technique suitable to study inhomogeneous fermion matter, such as the neutron star's crust, by means of molecular dynamics or Monte Carlo methods. The next step is to include the forces among the fermions to study the expected foam-like pasta structures.

\ack This work was supported by the Helmholtz International Center for FAIR (HIC for FAIR) within the framework of the LOEWE program launched by the State of Hesse (Germany).

\end{document}